\documentclass[aps,pre,preprint,showkeys,a4paper,amsmath,amssymb]{revtex4-1}
\usepackage[utf8]{inputenc}
\usepackage{booktabs}
\usepackage{graphicx,xcolor}
\usepackage{multirow}
\usepackage{dcolumn}
\usepackage{bm}
\usepackage{blindtext}
\usepackage[normalem]{ulem}

\begin{document}

\title{Boosting the performance of anomalous diffusion classifiers with the proper choice of features }

\author{Patrycja Kowalek}\email[Correspondence should be addressed to: ]{patrycja.kowalek@pwr.edu.com}
\author{Hanna Loch-Olszewska}
\author{\L{}ukasz \L{}aszczuk}
\author{Jaros\l{}aw Opa\l{}a}
\author{Janusz Szwabi\'nski}
\affiliation{Faculty of Pure and Applied Mathematics, Hugo Steinhaus Center, Wroc\l{}aw University of Science and Technology, 50-370 Wroc\l{}aw, Poland}

\begin{abstract}

Understanding and identifying different types of single molecules' diffusion that occur in a broad range of systems (including living matter) is extremely important, as it can provide information on the physical and chemical characteristics of particles' surroundings. In recent years, an ever-growing number of methods have been proposed to overcome some of the limitations of the mean-squared displacements approach to tracer diffusion. In March 2020, the Anomalous Diffusion (AnDi) Challenge was launched by a community of international scientists to provide a framework for an objective comparison of the available methods for anomalous diffusion. In this paper, we introduce a feature-based machine learning method developed in response to Task 2 of the challenge, i.e. the classification of different types of diffusion.
We discuss two sets of attributes that may be used for the classification of single-particle tracking data. The first one 
was proposed as our contribution to the AnDi Challenge. The latter is the result of our attempt to improve the performance of the classifier after the deadline of the competition. Extreme gradient boosting was used as the classification model. Although the deep-learning approach constitutes the state-of-the-art technology for data classification in many domains, we deliberately decided to pick this traditional machine learning algorithm due to its superior interpretability. After the extension of the feature set our classifier achieved the accuracy of 0.83, which is comparable with the top methods based on neural networks. 

\end{abstract}

\keywords{single particle tracking, anomalous diffusion, time series classification, machine learning, feature-based approach}

\maketitle

\section{\label{sec:intro}Introduction}

Single-particle tracking (SPT) is a popular method for observing the molecular dynamics in a wide range of materials, including the living cells~\cite{MAN15,SHE17}. Typically, an SPT experiment results in trajectories of tracers, i.e. time series of their consecutive positions, the analysis of which allows one to extract properties of the molecules and their environment (e.g. density of obstacles, velocity). 

The key first step in the analysis of SPT data is to connect the observed motion of particles with the available models of diffusion~\cite{Metzler2014Anomalous}, since it already sheds light on the mechanical properties of the particles' surrounding~\cite{mahowald2009impact}. The most common approach to characterize the diffusion of particles is the mean-squared displacement (MSD)~\cite{Metzler2014Anomalous,einstein1956investigations,Lemons1997}, which measures their deviations with respect to a reference position in time. A MSD growing linearly in time corresponds to Brownian motion (a.k.a. normal diffusion), which describes the movement of a microscopical particle as a consequence of thermal forces. Any deviations from that linear behavior are called anomalous diffusion~\cite{klages2008anomalous}, which can occur in physical, biological, or geological systems. Those deviations are usually represented by a power-law scaling of MSD (i.e. $MSD(t)\propto t^\alpha$)~\cite{Metzler2014Anomalous,einstein1956investigations,Lemons1997}, with $\alpha$ being the anomalous diffusion exponent. A sublinear MSD ($\alpha<1$) corresponds to subdiffusion. It indicates trapped particles~\cite{METZLER2000Therandomwalk,Hoze2012Heterogeneity}, particles which hit upon obstacles~\cite{Saxton1994-qu, Berry2014} or particles moderated by viscoelastic properties of the environment~\cite{WEISS2004Anomalous}. Subdiffusion can take place in crowded cellular fluids and membranes~\cite{Hfling2013}, and can be caused by crowding  and complex interactions with the cytoskeleton and macromolecular complexes: in cytoplasm \cite{golding2006physical,tolic2004anomalous,jeon2011vivo}, the nucleus \cite{bronstein2009transient} and the plasma membrane \cite{weigel2013quantifying,heinemann2013lateral,torreno2014enhanced}. A superlinear MSD ($\alpha>1$) is referred to as superdiffusion. In this case, the movement of particles is usually faster than in normal diffusion and frequently directed to a specific location. Examples of superdiffusion include L\'{e}vy flights \cite{bouchaud1990anomalous,ghosh2016anomalous}, active cytoplasmic flows and transport mediated by molecular motors~\cite{bursac2007cytoskeleton,kahana2008active,bursac2007cytoskeleton}, particle motion in random potentials or in turbulent flows \cite{richardson1926atmospheric,shlesinger1987levy}, foraging by animals and humans \cite{Viswanathan2009}, and many others~\cite{Zaburdaev2015}.

Despite its simplicity, the MSD-based analysis of SPT trajectories is truly challenging. The experimental trajectories are very often too short to extract meaningful information from the MSD. Moreover, the finite precision of tracers' localization adds a term to the MSD, which can limit the interpretation of data~\cite{MIC10,SAX1997,Kepten2015}. Consequently, several other approaches going beyond MSD  have been introduced. For instance, the radius of gyration~\cite{SAX93}, the velocity autocorrelation function~\cite{GRE13,FUL17}, the time-dependent directional persistence of trajectories~\cite{RAU07}, the distribution of directional changes~\cite{BUR13}, the mean maximum excursion method~\cite{TEJ10}, the fractionally integrated moving average (FIMA) framework~\cite{BUR15} may efficiently replace the MSD estimator for classification purposes. The full distribution of displacements may be fitted
to a mixed model in order to extract differences in diffusive
behavior between subsets of particle ensembles~\cite{SCH97}. Heterogeneity within single trajectories can be checked with Hidden Markov Models~\cite{DAS09,SLA15}. 

In the last few years, machine learning (ML) has become an interesting alternative for the analysis of anomalous diffusion. This approach is very appealing because in contrast to the analytical methods it does not require explicit rules for data processing. Instead, a ML algorithm can learn those rules directly from a series of data. ML is already known to excel in different domains including computer vision~\cite{KAR14}, speech recognition~\cite{DEN13a} and natural language processing~\cite{COL08}. First attempts to apply them to SPT data turned out to be very promising, even though the characterization of diffusion remains very challenging. Bayesian approach~\cite{MON12, THA18,Cherstvy2019}, random forests~\cite{WAG17, kowalek2019, MUN20,janczura2020,loch2020}, gradient boosting~\cite{kowalek2019, MUN20,janczura2020,loch2020}, neural networks \cite{DOS16}, and deep neural networks \cite{kowalek2019, BO19, GRA19, Gentili2021,Gajowczyk2021} have been used to either classify the trajectories or to extract quantitative information about them.

The increasing number of methods, both analytical and ML-based, and the lack of an objective comparison between them have resulted in the launch of the Anomalous Diffusion (AnDi) challenge by a team of international scientists~\cite{andi2020}. The goal of the challenge was at least two-fold: to provide a framework to benchmark the existing and new methods for the analysis of anomalous diffusion and to spur the invention of new high-performance methods. The challenge itself was divided into three main tasks: (1) inference of the anomalous diffusion exponent $\alpha$ from the trajectories provided by the organizers, (2) classification of trajectories into 5 different models of diffusion, and (3) detection of points within single trajectories, at which the anomalous exponent $\alpha$ and/or the diffusion coefficient $D$ change. From the results of the challenge, it follows that the ML methods based on neural networks outperform the more traditional statistics-based approaches~\cite{munozgil2021objective}. However, it should be emphasized here that the latter usually offer a deeper insight into the mechanisms of classification.

The main obstacle limiting the deployment of ML to the trajectory analysis is the availability of decent training data. The experimental trajectories are not provable (otherwise we would not need any new method of analysis). As a consequence, synthetic trajectories generated by computer simulations of various diffusion models are typically used as training data. The organizers of the AnDi challenge have identified five models of particular importance for the interpretation of experimental results: continuous time random walk (CTRW)~\cite{Scher1975}, fractional Brownian motion (FBM) \cite{Mandelbrot1968}, L\'{e}vy walk (LW)  \cite{shlesinger1987levy,Klafter1994,Klafter1996}, annealed transient time
motion (ATTM) \cite{Massignan2014} , and scaled Brownian motion (SBM) \cite{Lim2002}. To ensure that all participants of the challenge use the same data for inventing/improving their methods, a Python package with the implementation of the models has been provided by the organizers.

This paper describes the details of the method we proposed in response to the AnDi challenge. We focus on the classification of 2D trajectories, i.e. a part of the Task 2 within the challenge.
In our recent researches~\cite{kowalek2019,janczura2020,loch2020,Gajowczyk2021}, we already investigated the applicability of both feature-based models and deep learning techniques to anomalous diffusion. The first approach requires a set of human-engineered features to characterize each trajectory. Vectors of these features are then used as input for a classification model. Deep learning methods, in contrast, extract significant features from raw data on their own, without any effort from a human expert. The deep learning approach yielded slightly better performance than the feature-based ones, in line with the general findings of the AnDi challenge, but we decided to further elaborate on the latter due to its superior interpretability. Our method turned out to be inferior to the winning teams, but still offered a reasonable performance~\cite{munozgil2021objective}. Moreover, further investigation of the method after the  challenge allowed us to significantly improve its accuracy by adding some new features and to achieve performance similar to the best methods (0.83 accuracy in Task 2 versus 0.88 accuracy of the winners).

The paper is divided as follows.  Diffusion models used for training and the classification method are introduced in Sec.~\ref{sec:mod}. Sec.~\ref{sec:data} briefly summarizes the dataset used for training. Results of our analysis are presented in Sec.~\ref{sec:results}. Finally, some concluding remarks are given.

\section{\label{sec:mod}Models and Methods}

\subsection{Anomalous exponent $\alpha$}

The most popular method of deducing the particles' type of motion from their trajectories is based on the mean-squared displacement (MSD)~\cite{QIA91,Metzler2014Anomalous,einstein1956investigations,Lemons1997}, defined as
\begin{equation}
MSD(t) = \mathrm{E}\left( \Vert X_{t+t_0} -X_{t_0}\Vert^2 \right),
\label{eq:msd1}
\end{equation}
where $(X_t)_{t>0}$ is a particle's trajectory, $\Vert \cdot \Vert$ stands for the  Euclidean distance and $\mathrm{E}$ is the ensemble average. Since in many experiments only a limited number of trajectories is observed, the time-averaged MSD (TAMSD) calculated from a single trajectory is usually used as the estimator of MSD,
\begin{equation}
\widehat{MSD}(n\Delta t) =\frac{1}{N-n+1}\sum_{i=0}^{N-n} \Vert X_{t_{i+n}}-X_{t_i} \Vert^2.
\label{eq:tamsd}
\end{equation} 
The trajectory consists of $N$ consecutive two dimensional positions $X_i =(x_i,y_i)$ ($i=0,\dots,N$) recorded with a constant time interval $\Delta t$.  The quantity $n$ is the time lag between the initial and the final positions of the particle.  If the underlying process is ergodic and has stationary increments, TAMSD converges to the theoretical MSD~\cite{WEI11}.

Anomalous diffusion refers to a broad class of processes that deviate from the standard Brownian motion. Those deviations display an asymptotic power-law dependence of the MSD on the time lag,

\begin{equation}
\widehat{MSD}(n \Delta t) \sim  K_\alpha t^\alpha,\label{eq:msd}
\end{equation}
where $K_\alpha$ is the generalized diffusion coefficient and $\alpha$ is the anomalous exponent. As already mentioned in the introduction, the value of the latter may be used to discriminate the trajectories into normal diffusion ($\alpha=1$), subdiffusion ($\alpha<1$), and superdiffusion ($\alpha>1)$.

\subsection{Stochastic models of diffusion}

\begin{figure}
\includegraphics[scale=0.43]{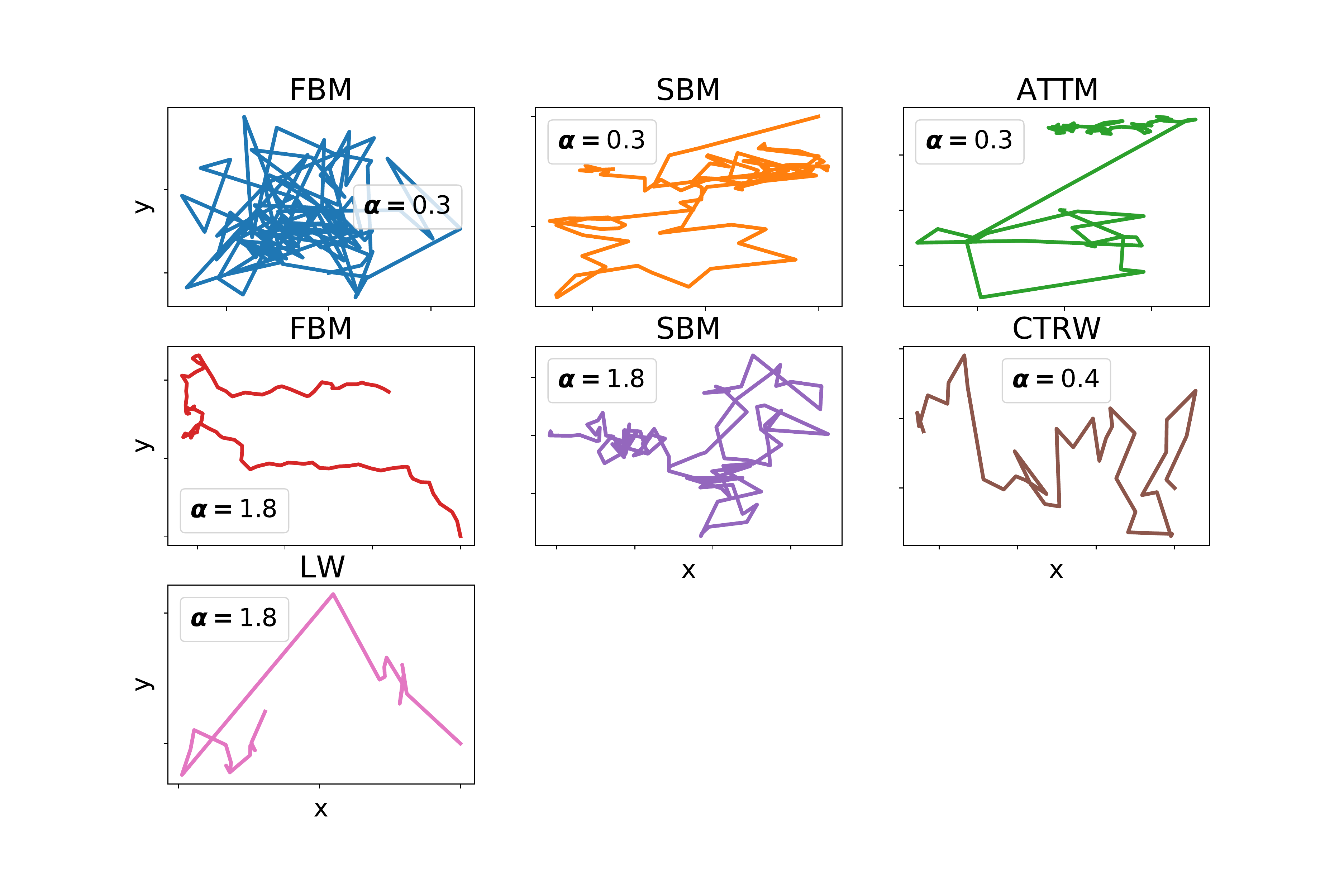}
\caption{Examples of trajectories generated with different anomalous exponents by the five models of diffusion: continuous time random walk (CTRW), fractional Brownian motion (FBM), L\'{e}vy walk (LW), annealed transient time motion (ATTM), and scaled Brownian motion (SBM). \label{fig:models}}
\end{figure}

Particles' movements that exhibit deviations from the linear behavior of MSD may  be described by multiple models, depending on some specific properties of the corresponding trajectories. Five of such processes have been identified within the AnDi challenge as crucial for the interpretation of the experimental data~\cite{andi2020,munozgil2021objective}: continuous time random walk (CTRW), fractional Brownian motion (FBM), L\'{e}vy walk (LW) , annealed transient time
motion (ATTM) and scaled Brownian motion (SBM) (see Fig.~\ref{fig:models} for example trajectories). In this section, the main characteristics of those models are briefly summarized.

\subsubsection{Continuous time random walk}

CTRW defines a large family of random walks with arbitrary displacement density for which the time spent between subsequent steps (i.e. the waiting time) is a stochastic variable~\cite{Scher1975}. Here, we will consider a particular instance of CTRW with the waiting times sampled from a power-law distribution $\psi(t) \sim t^{-\sigma}$, and the displacements drawn from Gaussian distribution with mean zero and variance $D$, $N(0,\sqrt{D})$. In this case, we have $\alpha=\sigma-1$.

\subsubsection{Fractional Brownian motion}

FBM~\cite{munozgil2021objective} is the solution of the stochastic differential equation
\begin{equation}
dX_t^i = \sigma dB_t^{H,i},~~i=1,2.
\label{eq:fbm}
\end{equation}
Here, $\sigma >0$ is the scale coefficient, which relates to the diffusion coefficient $D$ via $\sigma=\sqrt{2D}$. $H \in (0,1)$ is the Hurst parameter and $B_t^H$ is a continuous-time, zero-mean Gaussian process starting at zero, with the following covariance function
\begin{equation}
\mathrm{E}\left( B^H_t B^H_s\right) = \frac{1}{2}\left( |t|^{2H}+ |s|^{2H} - |t-s|^{2H} \right).
\end{equation}
The value of $H$ determines the type of diffusion in the process. For $H<\frac{1}{2}$, FBM produces subdiffusion (regime with a negatively correlated noise). For $H>\frac{1}{2}$, FBM generates superdiffusive motion (positively correlated noise). It reduces to the normal diffusion at $H=\frac{1}{2}$.

\subsubsection{L\'{e}vy walk}

L\'{e}vy walk~\cite{Klafter1994,shlesinger1987levy,Klafter1996} is another model of a random walk, but in contrast to CTRW, the waiting times and step lengths are coupled. In this work, we assume that waiting times are power-law distributed (as in case of CTRW, $\psi(t) \sim t^{-\sigma}$) and that the displacements and waiting times are correlated in such a way that the probability to move a certain step $\Delta x$ at time $t$ and stop at the new position to
wait for a new random event is $\Psi(\delta x,t)=\frac{1}{2}\psi(t)\delta(|\delta x|-vt)$, where $v$ is the velocity. Then, one can show that the anomalous exponent $\alpha=2$ for $0<\sigma<1$ and $\alpha=3-\sigma$ for $1<\sigma<2$. We may also observe then that the distribution of displacements is not Gaussian.

\subsubsection{Annealed transient time motion}

ATTM  describes the motion of a Brownian particle with a time-varying diffusion coefficient~\cite{Massignan2014}. The tracer performs Brownian motion with a random diffusion coefficient $D_1$ for a random time $t_1$, then with $D_2$ for $t_2$ and so on. The diffusion coefficients are drawn from a power-law distribution $P(D)\sim D^{\sigma-1}$. If the random times are sampled from a distribution with the expected value $E[t|D]=D^{-\gamma}$ ($\sigma<\gamma<\sigma+1$), then we have $\alpha = \sigma/\gamma$. According to Ref.~\cite{munozgil2021objective} we will assume that $P(t|D)\propto \delta(t-D^{-\gamma})$.

\subsubsection{Scaled Brownian motion}

The scaled Brownian motion (SBM) is a process
described by the Langevin equation with a time-dependent diffusivity $K(t)$,
\begin{equation}
    \frac{dX(t)}{dt} = \sqrt{2K(t)}\xi(t),
\end{equation}
where $\xi(t)$ is white Gaussian noise. If the diffusivity has the form $K(t)=\alpha K_\alpha t^{\alpha -1}$, the MSD follows the power-law $MSD(t)\propto K_\alpha t^\alpha$. 

\subsection{Classification model}

In machine learning, all classification algorithms may be divided into two classes. The traditional machine learning is a set of statistical learning methods, which do not operate on raw data. Instead, each data sample is characterized by a vector of human-engineered features or attributes. Those vectors are then used as input for the classifier. The second class consists of deep learning methods that identify and extract features on their own. The representation of data is constructed automatically and there is no need for its complex preprocessing as in the case of the feature-based methods. 

Deep learning is nowadays the state-of-the-art technology for data classification in many domains and overshadows a little bit the qualities of the classical machine learning. However, the choice of a suitable classification method is usually more subtle than simply looking at its performance.  A lot of deep learning methods do indeed an excellent job with respect to the predictions but are extremely complex to interpret. To give an example, the ResNet18 architecture we considered in one of our previous attempts to trajectory classification~\cite{Gajowczyk2021} originally had $11,220,420$ parameters. We were able to reduce their number down to $399,556$ with a positive impact on accuracy. Although it was an impressive achievement, interpretation of all of those remaining parameters is almost impossible. Thus, if one wants to comprehend the decisions made by an ML algorithm, it could be tempted to go for feature-based methods and gain on interpretability at the expense of accuracy.

Being aware of this trade-off, we deliberately decided to follow the interpretability path in our contribution to the AnDi challenge and chose the extreme gradient boosting (XGB) algorithm with decision trees as base learners \cite{RAS15} as the classification method. We already used similar approach~\cite{kowalek2019,janczura2020,loch2020}, however for different sets of data.

XGB is an ensemble method, which combines multiple decision trees  to obtain better predictive performance. A single decision tree \cite{song2015decision} is fairly simple to build.  The original dataset is split into smaller subsets based on the values of a given
feature. The process is recursively repeated until the resulting subsets are homogeneous (all samples
from the same class) or further splitting does not improve the classification performance. A splitting
feature for each step is chosen according to similarity score measure~\cite{Chen2016}. 

Single decision trees are famous for their ease of interpretation. However, they are sensitive to even
small variations of data and prone to overfitting. That is why one rather uses forests (ensembles) of trees as reliable classifiers. In the case of the XGB algorithm, the trees are built in a stage-wise fashion. We start with a single tree trained on random subsets of data and features and then at every step, a new tree is added to the ensemble. This new tree learns from the mistakes committed by the previous trees. 

\subsection{Diffusion characteristics for AnDi challenge}
\label{sec:orig features}

The main challenge of feature-based classification methods is the proper choice of features that can differentiate between samples belonging to different classes. The impact of their choice on the final performance of the classifier was already discussed in Ref.~\cite{loch2020}.  In this work, we continue the search for a robust set of features for the classification of different diffusion modes.

In this section, we briefly introduce features that have been used for the AnDi challenge. Since we were not satisfied with the performance of the resulting classifier in Task 2~\cite{munozgil2021objective}, after the end of the challenge we further searched for features that would improve the classification results. Those additional features will be described in Sec.~\ref{sec:add features}. Both the original feature set for the challenge and its extension are listed in Table~\ref{tab:features}.

\begin{table}
    \centering
    \begin{tabular}{c|c}
    \hline\hline
    Original features     & Additional features \\
    \hline\hline
    Anomalous exponent & D'Agostino-Pearson test statistic\\
    \hline
    Diffusion coefficient  & 
    \begin{tabular}{c}Kolmogorov-Smirnov statistic\\ against $\chi^2$ distribution\end{tabular}
    \\
    \hline
    Asymmetry & Noah exponent \\
    \hline
    Efficiency & Moses exponent\\
    \hline
    Empirical velocity autocorrelation function
     &  Joseph exponent
    \\
    \hline
    Fractal dimension & Detrending moving average \\
    \hline
    Maximal excursion &Average moving window characteristics\\
    \hline
    Mean maximal excursion & Maximum standard deviation \\
    \hline
    Mean gaussianity &\\
    \hline
    Mean-squared displacement ratio &\\
    \hline
    Kurtosis &\\
    \hline
    Statistics based on $p$-variation &\\
    \hline
    Straightness &\\
    \hline
    Trappedness & \\
    \hline\hline
    \end{tabular}
    \caption{The features used to characterize the SPT trajectories. The original set of features for the AnDi challenge (left column, see Sec.~\ref{sec:orig features} for definitions) have been extended afterwards (right columns, see Sec.~\ref{sec:add features} for further details) to improve the performance of the classifier.\label{tab:features}}
\end{table}

\subsubsection{Anomalous exponent}

In our set of features, we included 4 estimates for the anomalous exponent $\alpha$ (see Eq.~(\ref{eq:msd})), calculated using the following methods:
\begin{itemize}
    \item the standard estimation, based on fitting the empirical time-averaged MSD (TAMSD) from Eq.~(\ref{eq:tamsd}) to Eq.~(\ref{eq:msd}), 
    \item 3 estimation methods proposed for the trajectories with noise~\cite{lanoiselee2018_optimal}, under the assumption that the noise is normally distributed with zero mean:
    \begin{itemize}
        \item using estimator
        \begin{equation}
            \hat{\alpha} = \frac{n_{max} \sum_{n=1}^{n_{max}} \ln (n) \ln (\widehat{MSD}(n\Delta t)) - \sum_{n=1}^{n_{max}} \ln (n)\left( \sum_{n=1}^{n_{max}} \ln (\widehat{MSD}(n\Delta t)) \right)}{n_{max} \sum_{n=1}^{n_{max}} \ln^2(n) - \left(\sum_{n=1}^{n_{max}}\ln(n)\right)^2},
        \end{equation}
            with $n_{max}$ equal to 0.1 times the trajectory length, rounded to nearest lower integer (but not less than 4),
        \item  simultaneous fitting of parameters $\hat{D}$, $\hat{\alpha}$ and $\hat{\sigma}$ in the equation 
        \begin{equation}
            \widehat{MSD}(n\Delta t) = 2 d \hat{D}(n\Delta t)^{\hat{\alpha}}+\hat{\sigma}^2,
        \end{equation}
        where  $d$ denotes dimension, $D$ is the  diffusion coefficient and  $\sigma^2$ - the variance of noise,
        \item  simultaneous fitting of parameters $\hat{D}$ and $\hat{\alpha}$  in the equation
        \begin{equation}
            \widehat{MSD}(n\Delta t) = 2 d \hat{D} (\Delta t)^{\hat{\alpha}}(n^{\hat{\alpha}}-1).
        \end{equation}
    \end{itemize}
\end{itemize}

\subsubsection{Diffusion coefficient}

We used the diffusion coefficient extracted from the fit of the empirical TA-MSD to Eq.~(\ref{eq:msd}).

\subsubsection{Asymmetry}
The asymmetry of a trajectory can detect directed motion. According to Saxton~\cite{SAX93}, it can be derived from the gyration tensor, which describes the second moments of positions of a particle. For a 2D random walk of $N$ steps, the tensor is given by
\begin{equation}
\mathbf{T} =\left(
\begin{array}{cc}
\frac{1}{N}\sum_{j=1}^N (x_j -\langle x \rangle)^2 & \frac{1}{N}\sum_{j=1}^N (x_j -\langle x \rangle)(y_j -\langle y \rangle) \\ 
\frac{1}{N}\sum_{j=1}^N (x_j -\langle x \rangle)(y_j -\langle y \rangle) & \frac{1}{N}\sum_{j=1}^N (y_j -\langle y \rangle)^2
\end{array} 
\right), \label{eq:tensor}
\end{equation} 
where $\langle x \rangle=(1/N)\sum_{j=1}^N x_j$ is the average of $x$ coordinates over all steps in the random walk. We define the asymmetry as~\cite{HEL07}
\begin{equation}
A=-\log \left(1 - \frac{(\lambda_1-\lambda_2)^2}{2(\lambda_1+\lambda_2)} \right),\label{eq:asymmetry}
\end{equation}
where $\lambda_1$ and $\lambda_2$ are the principle radii of gyration, i.e. the eigenvalues of the tensor $\mathbf{T}$.

\subsubsection{Efficiency}
Efficiency $E$ measures the linearity of a trajectory. It relates the net squared displacement of a particle to the sum of squared step lengths,
\begin{equation}
E = \frac{|X_{N-1}-X_0|^2}{(N-1)\sum_{i=1}^{N-1}|X_i -X_{i-1}|^2}. \label{eq:efficiency}
\end{equation} 
It may help to detect directed motion (superdiffusion).

\subsubsection{Empirical velocity autocorrelation function}

Empirical velocity autocorrelation function \cite{Weber2010} for lag $1$ and point $n$ is defined as:
\begin{equation}
    \chi_{n} = \frac{1}{N-1}\sum^{N-2}_{i=0} (X_{i+1+n}-X_{i+n}) (X_{i+1}-X_{i})
\end{equation}
It can be used to distinguish subdiffusion processes. In our model, we used $\chi_n$ for points $n=1$ and $n=2$.

\subsubsection{Fractal dimension}
Fractal dimension is a measure of the space-filling capacity of a pattern (a trajectory in our case). According to Katz and George~\cite{Katz1985}, the fractal dimension of a planar curve may be calculated as 
\begin{equation}
    D_f=\frac{\ln N}{\ln (NdL^{-1})},
\end{equation}
where $L$ is the total length of the trajectory, $N$ is the number of
steps, and $d$ is the largest distance between any two positions. It takes values around $1$ for straight trajectories (i.e. directed motion), around 2  for random ones (normal diffusion), and around 3 for constrained trajectories (subdiffusion).

\subsubsection{Maximal excursion}
Maximal excursion of the particle is given by the formula:
\begin{equation}
    ME=\frac{\max(X_{i+1}-X_{i})}{X_{N-1}-X_0}
\end{equation}
It should detect relatively long jumps (in comparison to the overall displacement).

\subsubsection{Mean maximal excursion}

According to Ref.~\cite{TEJ10}, the mean maximal excursion is usually a better observable than MSD to determine the anomalous exponent $\alpha$. Given the largest distance traveled by a particle,
\begin{equation}
    D_N = \max |X_i-X_0|,
\end{equation}
the mean maximal excursion is defined as its standardized value, i.e.:
\begin{equation}
    T_n=\frac{\max(|X_i-X_0|)}{\sqrt{\hat{\sigma}^2_N(t_N-t_0)}},
\end{equation}
Here, $\hat{\sigma}_N$ is a consistent estimator of the standard deviation of $D_N$,
\begin{equation}
    \hat{\sigma}^2_N=\frac{1}{2N\delta t}\sum^{N}_{j=1}||X_j-X_{j-1}||^2_2.
\end{equation}

\subsubsection{Mean gaussianity}
Gaussianity $g(n)$~\cite{Ernst2014} checks the Gaussian statistics of increments of a trajectory and is defined as
\begin{equation}
    g(n)=\frac{2<r_n^4>}{3<r_n^2>^2},
\end{equation}
where $<r_n^k>$ denotes the $k$th moment of the trajectory at time lag $n$:
\begin{equation}
    <r_n^k>=\frac{1}{N-n}\sum^{N-n}_{i=1}|X_{i+n}-X_i|^k.
\end{equation}
Gaussianity for normal diffusion is equal to 0. The same result should be obtained for FBM, since its increments follow Gaussian distribution. Other types of motion should show deviations from zero.

Instead of looking at gaussianities at single time lags, we will include the mean over all lags as one of the features:
\begin{equation}
    \langle g\rangle= \frac{1}{N}\sum^{N}_{i=1}g(n).
\end{equation}

\subsubsection{Mean-squared displacement ratio}
MSD ratio gives information about the shape of the corresponding MSD curve. We will define it as
\begin{equation}
MSDR(n_1,n_2) = \frac{\langle r_{n_1}^2 \rangle}{\langle r_{n_2}^2 \rangle} - \frac{n_1}{n_2},\label{eq:msdr}
\end{equation}
where $n_1<n_2$. $MSDR=0$ is zero for normal diffusion ($\alpha=1$). We should get $MSDR\leq0$ for sub- and $MSDR\geq0$ for superdiffusion. We simply took $n_2=n_1+\Delta t$ and calculate an averaged ratio for every trajectory.

\subsubsection{Kurtosis}

Kurtosis gives insight into the asymmetry and peakedness of the distribution of points within a trajectory~\cite{HEL07}. To calculate it, the position vectors $X_i$ are projected onto the dominant eigenvector  $\vec{r}$ of the gyration tensor~(\ref{eq:tensor}),
\begin{equation}
x_i^p = X_i\cdot\vec{r}.
\end{equation}
Kurtosis is then defined as the fourth moment of  $x_i^p$,
\begin{equation}
K=\frac{1}{N}\sum_{i=1}^N \frac{(x_i^p -\bar{x}^p)^4}{\sigma^4_{x^p}},\label{eq:kurtosis}
\end{equation}
with $\bar{x}^p$ being the mean projected position and $\sigma_{x^p}$ - the standard deviation of $x^p$.

\subsubsection{Statistics based on $p$-variation}
The empirical $p$-variation is given by the formula \cite{BUR10,MAG09}:
\begin{equation}
V_m^{(p)}=\sum_{i=0}^{\frac{N}{m}-1}|X_{(i+1)m}-X_{im}|^p.
\end{equation}
These statistics can be used to detect the fractional L\'{e}vy stable motion (including FBM). We defined  6 features based on $V_m^{(p)}$:
\begin{itemize}
    \item power $\gamma^p$ fitted to $p$-variation for lags 1 to 5~\cite{BUR10}, 
    \item statistics $P$ used in Ref.~\cite{loch2020}, based on the monotonicity changes of $V_m^{(p)}$ as a function of $m$:
    	\begin{equation}
		P = \left\{
		\begin{array}{rl}
		0 & \textrm{if $V_m^{(p)}$ does not change the monotonicity}, \\ 
		1 & \textrm{if $V_m^{(p)}$ is convex for the highest $p$ for which it is not monotonous},\\
		-1 & \textrm{if $V_m^{(p)}$ is concave for  the highest $p$ for which it is not monotonous}.
		\end{array} 
		\right.
	\end{equation}
\end{itemize}

\subsubsection{Straightness}
Straightness $S$ measures the average direction change between subsequent steps. It relates the net displacement of a particle to the sum of all step lengths,
\begin{equation}\label{eq:straightness}
S = \frac{|X_{N-1}-X_0|}{\sum_{i=1}^{N-1}|X_i -X_{i-1}|}. 
\end{equation}

\subsubsection{Trappedness}

With trappedness we will refer to the probability that a diffusing particle is trapped in a bounded region with radius $r_0$. A comparison of analytical and Monte Carlo results for confined diffusion allowed Saxton~\cite{SAX93} to estimate this probability with
\begin{equation}
P(D,t,r_0) = 1-\exp\left( 0.2048 -0.25117\left( \frac{Dt}{r_0^2} \right)  \right).\label{eq:trappedness}
\end{equation}
Here, $r_0$ is approximated by half of the maximum distance between any two positions along a given trajectory. $D$ in Eq.~\eqref{eq:trappedness} is estimated by fitting the first two points of the MSD curve (i.e. it is the so called short-time diffusion coefficient).

\subsection{Additional diffusion characteristics}
\label{sec:add features}

In order to improve the performance of the original classifier, we searched for further feature candidates after the AnDi challenge. The additional features extending the basic set are described below.

\subsubsection{D’Agostino-Pearson test statistic}

D’Agostino-Pearson $\kappa^2$ test statistic~\cite{Agostino,AgostinoPearson} is a goodness-of-fit measure that aims to establish whether or not a given sample comes from a normally distributed sample. It is defined as
\begin{equation}
    \kappa^2 = Z_1(g_1) + Z_2(K),
\end{equation}
where $K$ is the sample kurtosis given by Eq.~(\ref{eq:kurtosis}) and $g_1=m_3/m_2^{3/2}$ is the sample skewness with $m_j$ being the $j$th sample central moment. The transformations $Z_1$ and $Z_2$ should bring the distributions of the skewness and kurtosis as close to the standard normal as possible. Their definitions may be found elsewhere~\cite{Agostino,AgostinoPearson}. This feature should help to distinguish ATTM and SBM from other motions.

\subsubsection{Kolmogorov-Smirnov statistic against $\chi^2$ distribution}

The Kolmogorov-Smirnov statistic quantifies the distance between the empirical distribution function of the sample $F_n(X)$ and the cumulative distribution function $G_n(X)$ of a reference distribution,
\begin{equation}
   D_n = \sup_X \vert F_n(X) - G_n(X) \vert .
\end{equation}
Here, $n$ is the number of observations (i.e. the length of a trajectory). The value of this statistic for the empirical distribution of squared increments of a trajectory against the sampled $\chi^2$ distribution has been taken as the next feature. The rationale of such choice is that for a Gaussian trajectory the theoretical distribution of squared increments is the mentioned $\chi^2$ distribution.

\subsubsection{Noah, Moses, and Joseph exponents}

For processes with stationary increments, there are in principle two mechanisms that violate the Gaussian central limit theorem and produce anomalous scaling of MSD: long-time increment correlations or a flat-tailed increment distribution (in the latter case the second moment is divergent)~\cite{Aghion2021}. These mechanisms are referred to as the Joseph and Noah effects, respectively. FBM is the prototypical process that exhibits the Joseph effect. LW on the other hand is an example of a process with the Noah effect. An anomalous scaling can also be induced by a non-stationary increment distribution~\cite{Aghion2021}. In this case, we deal with the Moses effect. It should help to handle ATTM and SBM trajectories.

All three effects may be quantified by exponents, which will be used as features in our extended set of attributes. Given a stochastic process $X_t$ and the corresponding increment process $\delta_t(\tau)=X_{t+\tau}-X_t$, the Joseph, Moses and Noah exponents are defined as follows.

\begin{enumerate}
    \item Joseph exponent $J$ is estimated from the ensemble average of the rescaled range statistic:
$$E\left[\frac{\max_{1\leq  s\leq  t}[X_s-\frac{s}{t}X_t]-\min_{1\leq  s\leq  t}[X_s-\frac{s}{t}X_t]}{\sigma_t}\right]\sim t^J,$$    
    
$$E\left[\frac{R_t}{S_t}\right]\sim t^J,$$ where $R_t$ is calculated as $R_t=\max_{1\leq  s\leq  t}[X_s-\frac{s}{t}X_t]-\min_{1\leq  s\leq  t}[X_s-\frac{s}{t}X_t]$ and $S_t$ is standard deviation of process $X_t$.
\item Moses exponent $M$ is determined from the scaling
of the ensemble probability distribution of the sum of the
absolute value of increments, which can be estimated by
the scaling of the median of the probability distribution
of $Y_t=\sum^{t-1}_{s=0}|\delta_s|$:
$$E[|\delta_t-E(\delta_t)|]\sim t^{M-\frac{1}{2}}$$
\item Noah exponent $L$ is extracted from the scaling of
the ensemble probability distribution of the sum of increment
squares, which can be estimated by the scaling of the median
of the probability distribution of $Z_t=\sum^{t-1}_{s=0}\delta_s^2$:
$$m[Z_t]\sim t^{2L+2M-1}$$
\end{enumerate}

\subsubsection{Detrending moving average}
The detrending moving average (DMA) statistic~\cite{sikora2018_statistical,balcerek2021} is given by
\begin{equation}
    DMA(\tau) = \frac{1}{N-\tau+1} \sum_{i=\tau}^{N} \left( X_i - \overline{X}^{\tau}_i \right)^2,
\end{equation}
for $\tau=1,2,...$, where $\overline{X}^{\tau}_i$ is a moving average of $\tau$ observations, i.e. $\overline{X}^{\tau}_i = \frac{1}{\tau+1} \sum_{j=0}^{\tau} X_{i-j}$.

As mentioned in Ref.~\cite{balcerek2021}, the DMA-based statistical tests can help in the detection of the scaled Brownian motion. In our model, we used two values of DMA for each trajectory as input features, namely $DMA(1)$ and $DMA(2)$.

\subsubsection{Average moving window characteristics}

As the moving average methods have already been successfully applied to many problems (see for example Ref.~\cite{hubicka2020} for change point detection), we decided to accommodate them in our feature set as well. We added eight features based on the formula
\begin{equation}
    MW = \frac{1}{2(N+1)} \sum_{t=0}^{N} \left| \textrm{sgn} \left(\overline{X}_{i+1}^{(m)}-\overline{X}_{i}^{(m)}\right) - \textrm{sgn}\left(\overline{X}_{i+1}^{(m)}-\overline{X}_{i}^{(m)}\right) \right|,
\end{equation}
where $\overline{X}^{(m)}$ denotes a statistic of the process calculated within the window of length $m$ and $\textrm{sgn}$ is the signum function. In particular, we used the mean and the standard deviation for $\overline{X}$ and calculated $MW$ with windows of  lengths $m=10$ and $m=20$ separately  for $x$ and $y$ coordinates.

\subsubsection{Maximum standard deviation}

The idea of the moving window helped us to introduce another two features based on the standard deviation $\sigma_m$ of the process calculated within a window of length $m$. They are given by 
\begin{equation}
    MXM = \frac{\max\left(\sigma_m(t)\right)}{\min\left(\sigma_m(t)\right)}
\end{equation}
and
\begin{equation}
    MXC = \frac{\max\left|\sigma_m(t+1)-\sigma_m(t)\right|}{\sigma},
\end{equation}
where $\sigma$ denotes the overall standard deviation of the sample. We used the window of length $m=3$ and calculated the features for both coordinates separately. They should improve the detection of ATTM type of movements.

\section{\label{sec:data}Data}

Extreme gradient boosting, i.e. the algorithm we decided to use, is an example of a supervised model. In other words, it requires a training dataset consisting of trajectories and the corresponding labels indicating their diffusion type (i.e. the diffusion model). 

We used a synthetic set of trajectories produced with the \texttt{andi-dataset} package~\cite{zendo_adni2020} to train the classifier. Once the trajectories were generated, a vector of features was calculated for each of them in the preprocessing phase. The resulting set consisting of vectors of attributes and their labels was used as input for the XGB model.

We generated $3\times 10^5$ trajectories for each diffusion model. Those trajectories were evenly distributed over 20 ranges of lengths, from very short ones ($n\in \langle10,50))$ to really long ones ($n\in(951,1000\rangle$). The whole dataset was divided into training, test, and validation subsets in the proportion of 0.7:0.15:0.15, respectively. The classifier is initially fit on the training data. The validation set provides an unbiased evaluation of a model while tuning its hyperparameters. Test data is used to provide an unbiased assessment of the final model. Stratified sampling was used to ensure a proper balance of data in the subsets. It should be noted that in this article, the dataset is significantly larger than the one used for the AnDi challenge~\cite{andi2020} (only $7\times 10^4$ trajectories). 

The parameters used to generate the trajectories are shown in Table~\ref{tab:param}. Their ranges have been constrained by the organizers of the challenge. The range of diffusion exponents prevents stationary paths. All trajectories were standardized so that the distribution of displacements over the unit time is characterized by the standard deviation equal to 1. 

To better simulate experimental endeavors, each trajectory was contaminated with a finite localization error. For that purpose, a random number from a normal distribution $N(0,\sigma_{\text{noise}})$ was added to each trajectory coordinate. 

\begin{center}
\begin{table}
\begin{tabular}{c|c|c}
\hline \hline
Parameter & Meaning & Parameter ranges   \\ 
\hline 
$\alpha$ & diffusion exponent & $\langle 0.05, 2\rangle$  \\ 
\hline 
$n$     & length of a trajectory & $\langle 10,1000)$   \\ 
\hline 
$\sigma_{\text{noise}}$   & standard deviation (level) of noise & $ \{0.1,0.5,1\}$   \\ 
\hline\hline 
\end{tabular} 
\caption{\label{tab:param} Parameters of the diffusion models used to generate the synthetic trajectories.}
\end{table}
\end{center}

\section{\label{sec:results}Results}

We used the implementation of the XGB from the \texttt{xgboost} module~\cite{Chen2016}. All computations were carried out on a cluster of 24 CPUs with a total memory of 25 GB. A randomized search strategy over the hyperparameter space of the classifier (the \texttt{RandomizedSearchCV} function in \texttt{scikit-learn} module~\cite{PED11}) has been deployed in order to optimize the model. The final values of the hyperparameters are summarized in Table~\ref{tab:hypparam}. The parameter \verb?min_child_weight? indicates the minimum of instance weight (corresponding to the number of instances) needed in a child node of the single decision tree.  If the tree partition step results in a leaf node with the sum of instance weight less than its value, the building process of the tree will give up further partitioning. \verb?max_depth? determines the maximum depth of a tree. The value of \texttt{gamma} indicates the minimum loss reduction to make a further split on a leaf node of the tree.  The evaluation metric for the validation step is specified by \verb?eval_metric?. Mean absolute error (\texttt{mae}) is used in our case. \texttt{eta} is the learning rate used to prevent overfitting of the classifier. And the \verb?base_score? is the global bias, i.e. the initial prediction score of all instances. 

The values of the hyperparameters were optimized for the original set of features. However, the same values of hyperparameters were also used with the extended set for comparison purposes.

\begin{center}
\begin{table}
\begin{tabular}{c|c}
\hline \hline
\texttt{Hyperparameter}  & Parameter value   \\ 
\hline 
\texttt{min\_child\_weight}  & $7$  \\ 
\hline 
\texttt{max\_depth}    & $10$   \\ 
\hline 
\texttt{gamma}     &  $4$   \\ 
\hline 
\texttt{eval\_metric}  &  mae   \\ 
\hline
\texttt{eta} & $0.3$\\
\hline
\texttt{base\_score} & $1$\\
\hline\hline 
\end{tabular} 
\caption{Hyperparameters of the final extreme gradient boosting models used for classification.\label{tab:hypparam}}
\end{table}
\end{center}

\subsection{Performance of the classifiers}

The performance of the classifiers in Task 2 of the AnDi challenge was evaluated with the $F_1$ score,
\begin{equation}
    F_1 = \frac{TP}{TP+0.5(FP+FN)}~\label{eq:f1}
\end{equation}
where $TP$, $FP$, and $FN$ are the true positive, false positive, and false negative rates, respectively. Since we are dealing here with a multilabel classification problem, a micro-averaged $F_1$ score was calculated to assess the aggregated contributions of all classes (i.e. the sums of all $TP$, $FP$ and $FN$ were first determined and then inserted into Eq.~\ref{eq:f1}. For balanced sets (same number of samples for each class), like those of the AnDi challenge, the micro-averaged $F_1$ score coincides with the accuracy of the classifier.

\begin{center}
\begin{table}
\begin{tabular}{c|c}
\hline \hline
Base  XGB model  & Extended XGB model   \\ 
\hline 
0.73  & 0.83  \\ 
\hline\hline 
\end{tabular} 
\caption{Micro-averaged $F_1$ score (i.e. accuracy) of both the original and the extended classifiers (see Sec.~\ref{sec:orig features} and \ref{sec:add features} for more details).\label{tab:results}}
\end{table}
\end{center}

Results for the performance of the classifiers are shown in Table~\ref{tab:results}.
As we can see, the model with the original set of features achieves an accuracy of 0.73. After adding the features described in Sec.~\ref{sec:add features} we observe a significant improvement of its performance. The accuracy of the extended version (0.83) is comparable with the best methods of the AnDi challenge~\cite{munozgil2021objective}.

\begin{figure}
\includegraphics[scale=0.45]{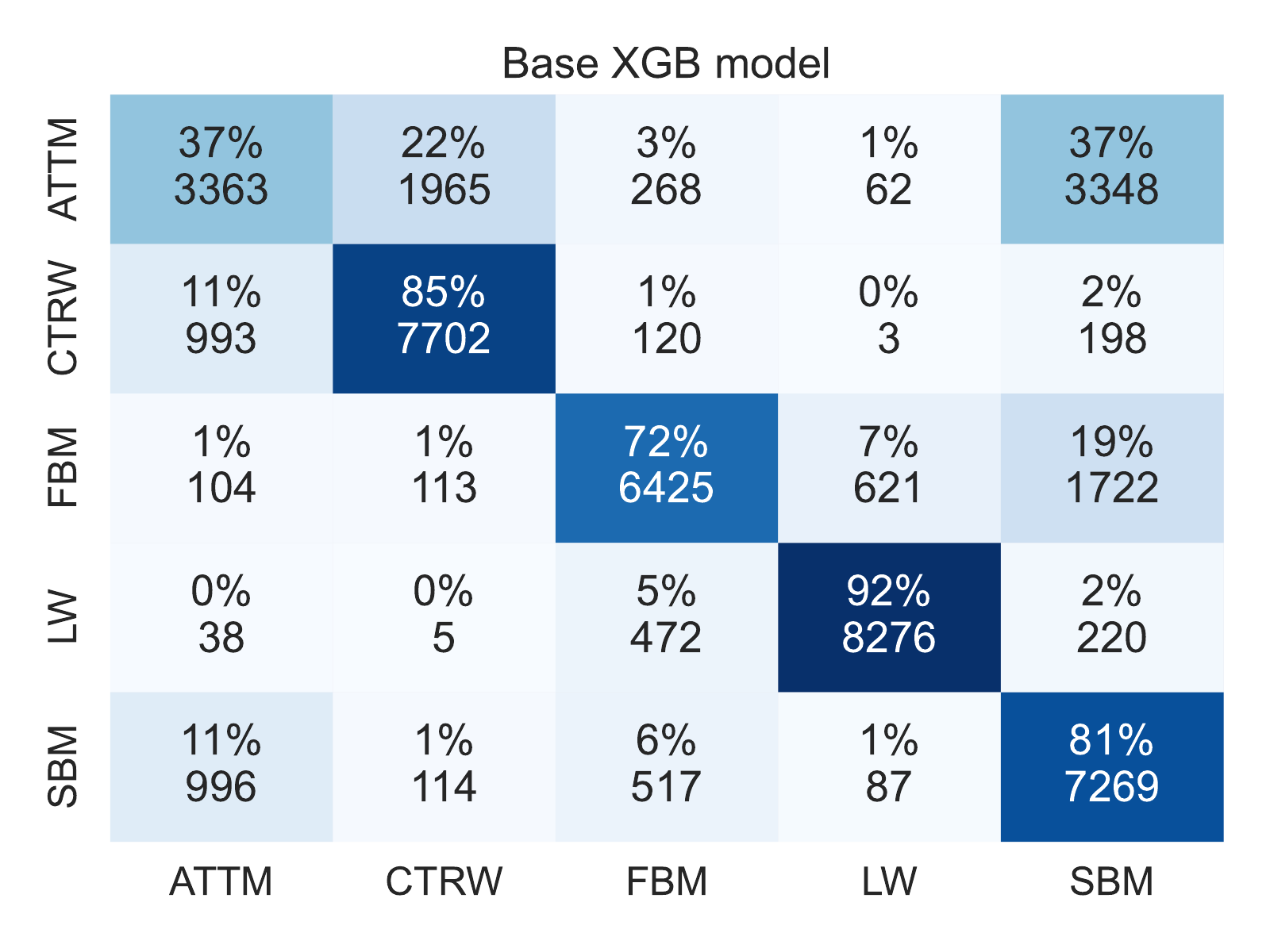}
\includegraphics[scale=0.45]{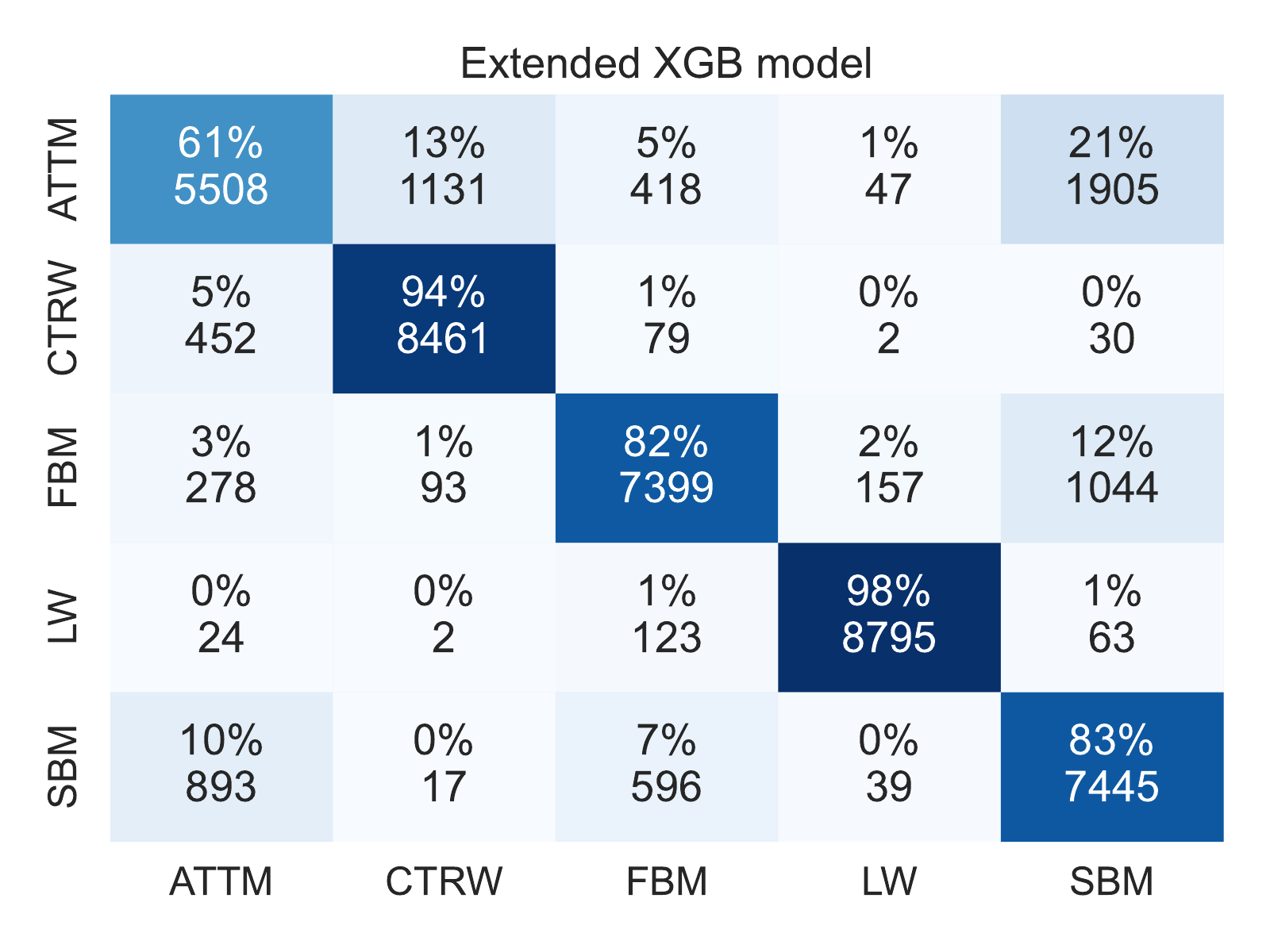}
\caption{Normalized confusion matrices of the models. Rows correspond to the true labels and columns to the predicted ones. \label{fig:cm1}}
\end{figure}

Inspection of the confusion matrices of the classifiers may give us additional insight into partial contributions of the overall performance. To recall, an element $c_{ij}$ of the confusion matrix indicates how many observations known to belong to class $i$ (true label) are predicted to be in class $j$ (predicted labels)~\cite{RAS15}. Normalized confusion matrices for both models are shown in Fig.~\ref{fig:cm1}.
The best performance of the base model is observed for L\'{e}vy walks. Among all LW samples in the test data, 92\% of them were correctly recognized. Most of the misclassified LW trajectories were assigned to FBM motion (5\%) and SBM (2\%). The accuracy of the classifier for CTRW (85\%) and SBM (81\%) is smaller, but still quite good. The incorrectly classified CTRW trajectories have been usually confused with ATTM. In the case of SBM, many trajectories have been assigned to ATTM (11\%) and FBM (6\%) classes. 

We have not expected such a moderate performance of the original classifier for FBM (72\%). In our earlier attempts, \cite{kowalek2019,janczura2020,loch2020} we usually got much higher performance for this type of motion. However, it should be stressed that previously we trained our classifiers on different set of  trajectories  than used for the AnDi challenge.
Moreover, we usually used longer trajectories. Thus, it seems that the features in the original set are lacking the power to differentiate FBM successfully from other types of motion considered in this paper, in particular from SBM (19\% of misclassifications) and LW (7\%).

Finally, it should be emphasized that the performance of the base classifier for ATTM is unsatisfactory (37\%). Most of the ATTM trajectories are confused with SBM (37\%) and CTRW (22\%). It is clear that the base model cannot separate the different origins of trajectories from models with time-varying diffusion coefficient $D$.

Adding the features from Sec.~\ref{sec:add features} to the input vectors has significantly improved the overall accuracy of the classifier. As can be seen in the right plot of Fig.~\ref{fig:cm1}, all of the partial contributions to the accuracy improved as well. Although the performance for ATTM remains weak (61\% only), it is much higher than in the previous case. As for the other classes, the modified classifier achieves very good accuracies for LW (98\%) and CTRW (94\%) and good ones for SBM (83\%) and FBM (82\%). The biggest challenge is still to distinguish ATTM and FBM from SBM trajectories.

\subsection{Impact of trajectory lengths}
\begin{figure}
\includegraphics[scale=0.45]{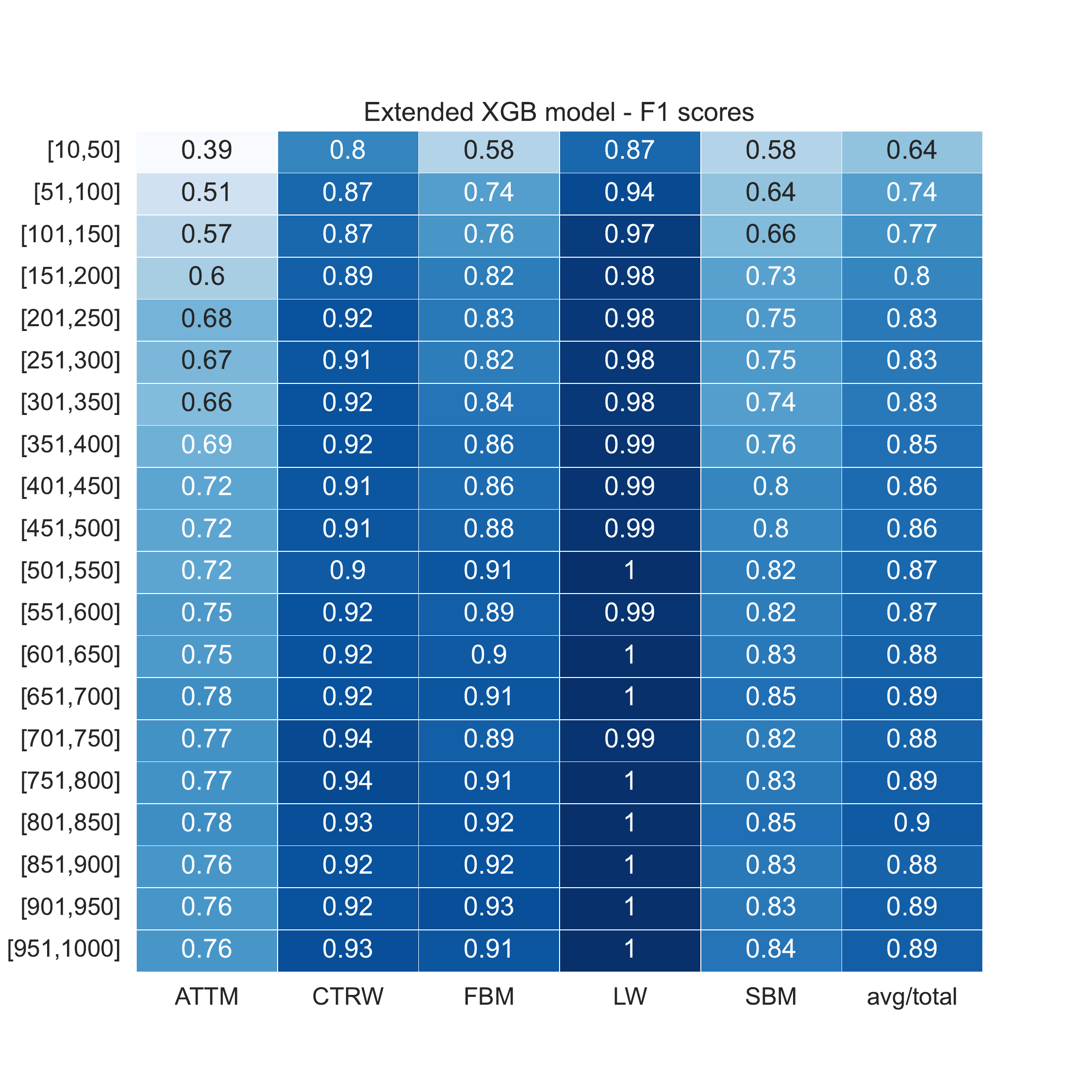}
\caption{$F_1$ scores as functions of different lengths of the trajectories. Columns represent $F_1$ score for each type of the motion, rows correspond to ranges of the lengths. \label{fig:f3}}
\end{figure}

In Fig.~\ref{fig:f3}, the $F_1$ scores for different ranges of the trajectory lengths in the case of the extended model are shown. Not surprisingly, the classifier struggles a lot with very short trajectories. In this regime, many of the paths are indistinguishable from normal diffusion. Moreover, in the case of ATTM and SBM models, the information contained in the short trajectories may be not enough for detecting the time-varying diffusion coefficient $D$. Interestingly, the accuracy of the FBM classification increases with the sample length to achieve around 90\% for trajectories longer than 500 steps.

It is worth to mention that the difficulties with very short trajectories are not specific to the extreme gradient boosting in particular or to feature-based methods in general. All methods presented in the challenge were afflicted by similar problems in this regime~\cite{munozgil2021objective}.  

\subsection{Impact of the level of noise}

\begin{figure}
\includegraphics[scale=0.45]{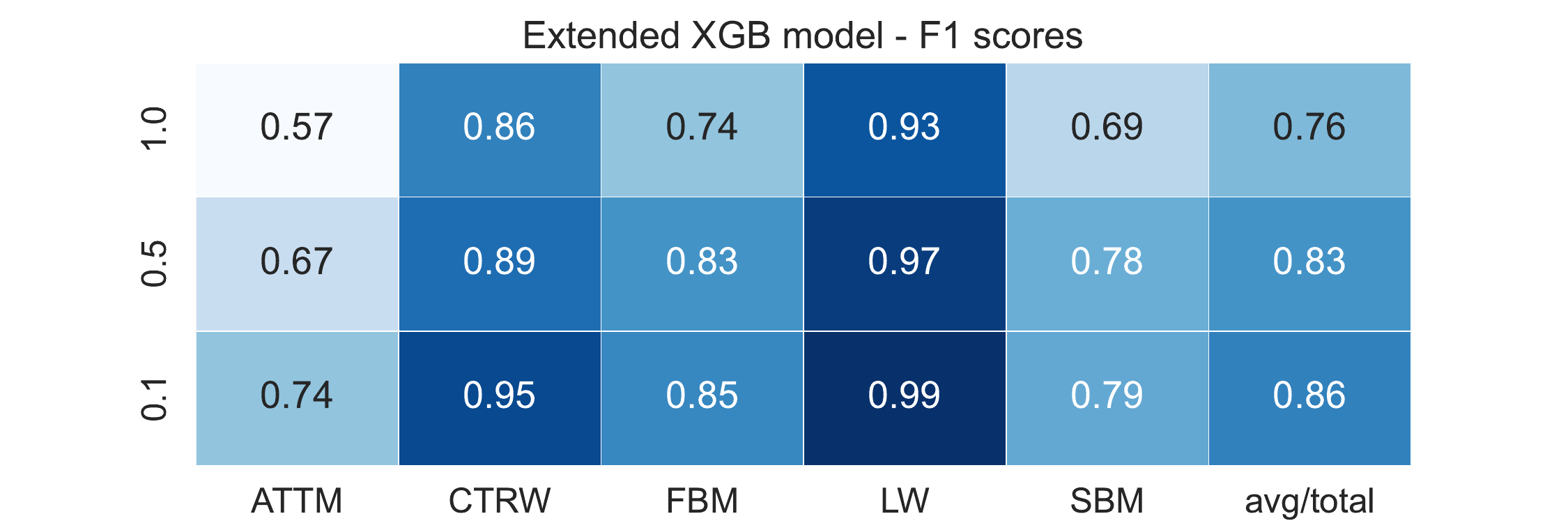}
\caption{ $F_1$ scores for different levels of trajectories' noise. Columns represent $F_1$ scores for each type of the motion, rows correspond to standard deviations of noise. \label{fig:f4} }
\end{figure}

In Fig.~\ref{fig:f4}, the $F_1$ scores for different levels of noise (i.e. different values of its standard deviation) are presented in case of our extended model. While the classifier performs quite well for the CTRW and LW even in the case of high noise levels, its accuracy significantly decreases for the remaining diffusion models. It seems that the classifier is not able to capture enough characteristics of ATTM, SBM and FBM types of movements from noise trajectories in order to successfully distinguish them from each other. Interestingly, in a series of computer experiments we observed that a classifier trained on very noisy data performs still reasonably well on a test set with low levels of noise. 

\subsection{Features importance}

One of the benefits of feature-based methods is the relative ease of their interpretation. In particular, it is possible to assign a score to each of the features indicating how important these features are to the model's predictions. This importance aids to extract relevant knowledge concerning the relationships contained in the data or learned by the model. But it can also help in the selection of the meaningful features for a simplified version of the classifier.

One of the popular feature assessment techniques is the permutation feature importance~\cite{RAS15}. It is defined as the decrease in a classifier's score when a single feature values are randomly shuffled. Since this procedure breaks the relationship between the feature and the target, the drop in the score indicates how much the classifier depends on the feature.  In Table~\ref{tab:permuatationfeatureimportance}, the 20 most important features in our model according to the permutation method are presented. It is worth to note that the features added in the extended version of the model play the key role among them. In particular, the average moving window (MW) features have the highest score, followed by the D'Agostino-Pearson test statistic $\kappa^2$.

\begin{center}
\begin{table}
\begin{tabular}{|c|c|c|}
\hline \hline
model characteristic  & mean feature importance & std. deviation   \\   \hline
\texttt{mw\_y\_mean10}                             & 0.1089                  & 0.0015                                     \\\hline
\texttt{mw\_x\_mean10}                             & 0.1013                  & 0.0008                                     \\\hline
\texttt{dagostino\_y}                              & 0.0779                  & 0.0016                                     \\\hline
\texttt{dagostino\_x}                              & 0.0727                  & 0.0010                                     \\\hline
\texttt{M}                                         & 0.0645                  & 0.0003                                     \\\hline
\texttt{mw\_y\_mean20}                             & 0.0516                  & 0.0009                                     \\\hline
\texttt{mw\_x\_mean20}                             & 0.0497                  & 0.0007                                     \\\hline
\texttt{max\_std\_y}                               & 0.0422                  & 0.0013                                     \\\hline
\texttt{max\_std\_x}                               & 0.0402                  & 0.0010                                     \\\hline
\texttt{alpha}                                     & 0.0295                  & 0.0008                                     \\\hline
\texttt{p\_var\_1}                                 & 0.0283                  & 0.0005                                     \\\hline
\texttt{fractal\_dimension}                        & 0.0272                  & 0.0007                                     \\\hline
\texttt{mean\_gaussianity}                         & 0.0240                  & 0.0004                                     \\\hline
\texttt{vac\_lag\_1}                               & 0.0211                  & 0.0013                                     \\\hline
\texttt{ksstat\_chi2}                              & 0.0194                  & 0.0011                                     \\\hline
\texttt{max\_std\_change\_x}                       & 0.0178                  & 0.0012                                     \\\hline
\texttt{max\_std\_change\_y}                      & 0.0111                  & 0.0007                                     \\\hline
\texttt{max\_ts}                                   & 0.0078                  & 0.0007                                     \\\hline
\texttt{L}                                         & 0.0074                  & 0.0005                                     \\\hline
\texttt{D}                                         & 0.0068                  & 0.0007                                    \\
\hline\hline 
\end{tabular} 
\caption{Permutation feature importance values for the XGB classifier.\label{tab:permuatationfeatureimportance}}
\end{table}
\end{center}

Since the permutation importance is known to have several drawbacks while working with correlated features~\cite{hooker_mentch_zhou_2021} and this is the case in our model (e.g. we have several features based on MSD), 
we decided to check the gain importance as well~\cite{XGB_doc}. In this approach, the importance of a feature represents the relative contribution of the feature to the model, calculated by taking each feature’s contribution for each tree in the model. A higher value of this metric when compared to other features implies it is more important for generating predictions. The top 20 features according to gain are listed in Table~\ref{tab:gainfeatureimportance}. Although the moving average window and D'Agostino-Pearson characteristics are still very important, the anomalous exponent $\alpha$ and the MSD ratio get now the highest scores. But again, many of the most important features were added in the extended version of the model. Thus it seems we made some reasonable choices to improve the classification accuracy.

\begin{center}
\begin{table}
\begin{tabular}{|c|c|}
\hline \hline
model characteristic  & Feature Importance   \\   \hline
\texttt{alpha}                                       & 177.31                             \\ \hline
\texttt{mean\_squared\_displacement\_ratio}          & 135.70                             \\ \hline
\texttt{mw\_x\_mean10}                               & 92.59                             \\ \hline
\texttt{mw\_y\_mean10}                               & 60.82                             \\ \hline
\texttt{dagostino\_y}                                & 60.21                             \\ \hline
\texttt{max\_ts}                                     & 48.25                             \\ \hline
\texttt{max\_std\_x}                                 & 39.82                             \\ \hline
\texttt{dagostino\_x}                                & 38.16                             \\ \hline
\texttt{max\_std\_y}                                 & 27.55                             \\ \hline
\texttt{M}                                           & 24.56                              \\ \hline
\texttt{fractal\_dimension}                          & 23.44                             \\ \hline
\texttt{p-variation}                                 & 20.20                             \\ \hline
\texttt{mean\_gaussianity}                           & 13.93                             \\ \hline
\texttt{mw\_y\_std10}                                & 13.27                             \\ \hline
\texttt{p\_var\_1}                                   & 13.02                             \\ \hline
\texttt{mw\_x\_std10}                                & 12.84                             \\ \hline
\texttt{max\_std\_change\_x}                         & 11.01                              \\ \hline
\texttt{L}                                           & 10.73                             \\ \hline
\texttt{max\_std\_change\_y}                         & 9.88                             \\ \hline
\texttt{mw\_x\_std20}                                & 9.75                             \\ 
\hline\hline 
\end{tabular} 
\caption{Gain feature importances for the XGB classifier.\label{tab:gainfeatureimportance}}
\end{table}
\end{center}

Last but not least, to further elaborate on that issue, we will employ another well known technique - the SHAP values based on feature attribution, i.e. the impact of having a certain value of a given feature in comparison to the prediction we'd make if that feature took some baseline value~\cite{lundberg2017unified,lundberg2018Consistent}. Compared the the previous methods, SHAP values may be used to explain individual predictions. Moreover, they offer a more consistent approach to assess the overall contribution of features to the final outcome of the model.

In Table~\ref{tab:featureimportance}, SHAP values for the most important features in different diffusion models are shown, in the case of the extended model. It should be noted that for every single diffusion model, the additional features turned out to be the most important ones, in agreement with the previous methods. In other words, the additional features prove to be an excellent choice. This also shows how critical the choice of features can be for a classifier (see Ref.~\cite{loch2020} for further details).

The Moses exponent $M$, which relates to non-stationary increments, turned out to be the essential attribute for distinguishing ATTM and SBM from the rest of models; it seems to be quite important for FBM as well. However, in the latter case, the anomalous exponent is the most meaningful feature, followed by the D’Agostino-Pearson test statistic $\kappa^2$. Characteristics based on the average moving window ($MW$) have the highest importance for CTRW. And finally, the maximum standard deviation feature is the top one for LW.

\begin{figure}
\centering
{\includegraphics[width=0.49\textwidth]{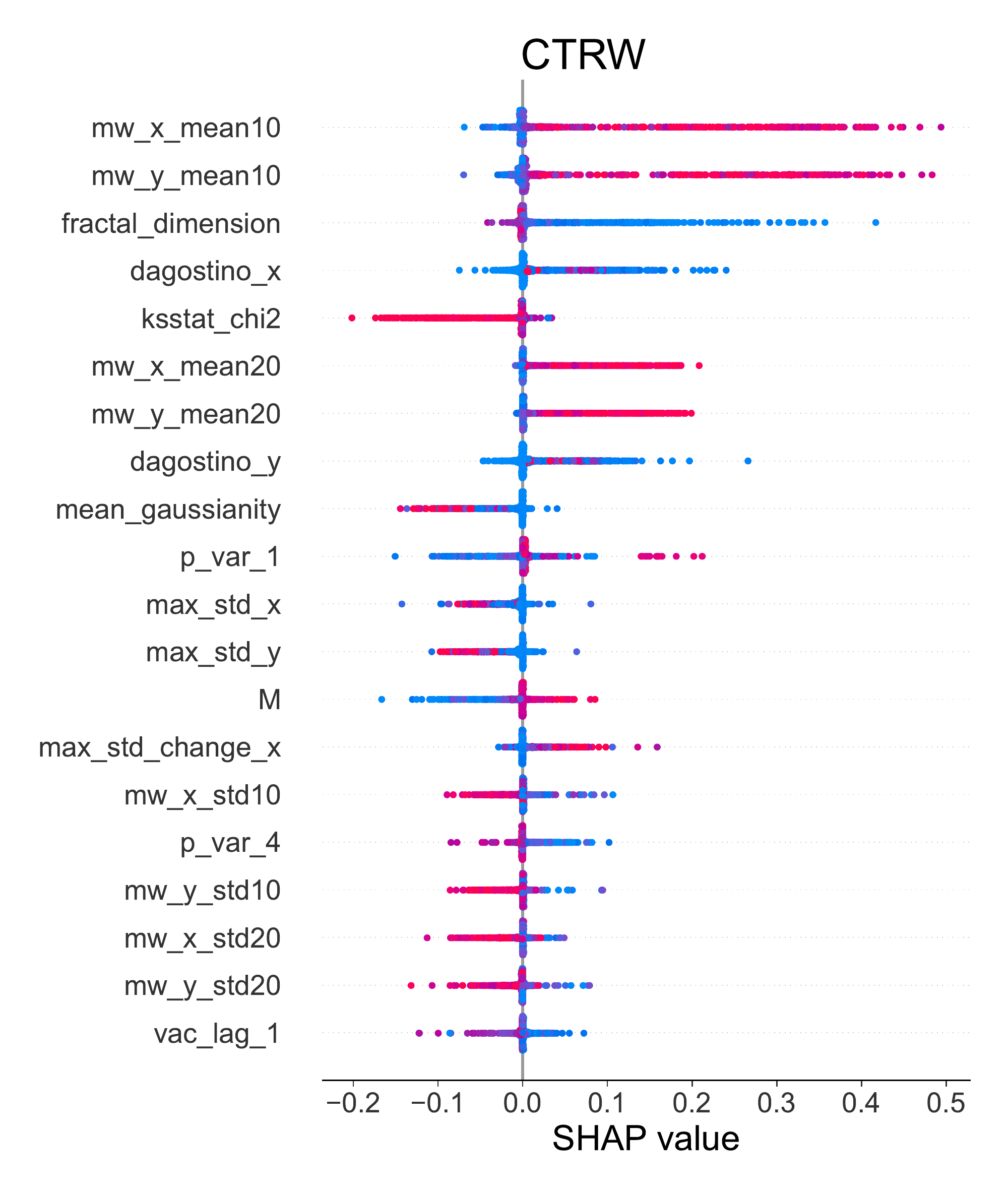}} 
{\includegraphics[width=0.49\textwidth]{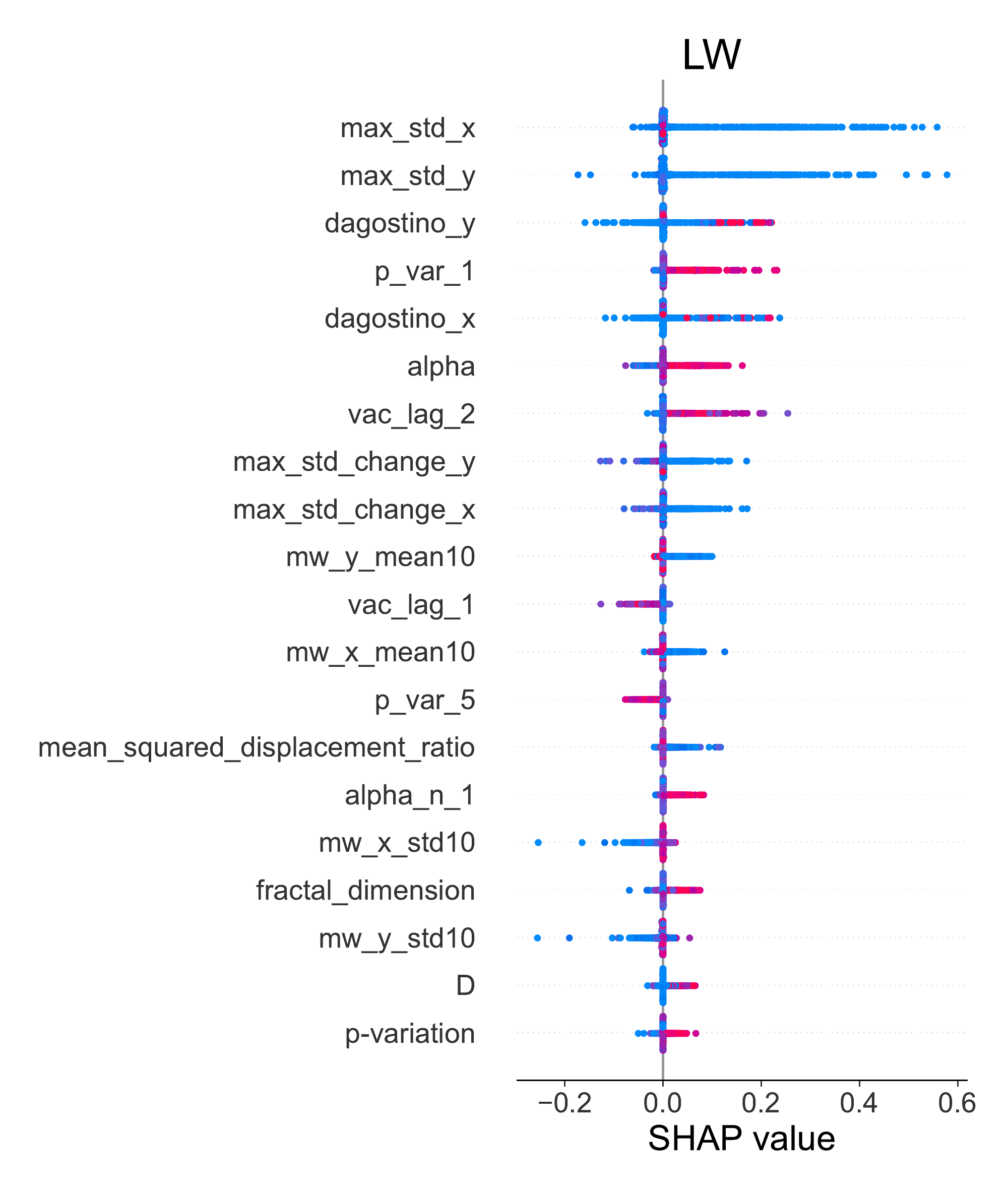}} 
\caption{SHAP values for 20 most relevant features in case of CTRW and LW. Dots represent the value for a single explanation (trajectory). The characteristics on the vertical axis are sorted by their mean importance. The SHAP values on the horizontal axis reflect the
positive or negative influence of each factor on the model prediction. A positive SHAP value improves the likelihood that the sample will be classified as a specific model. A longer tail indicates a greater impact on the prediction. 
The dependent variable Y is positively correlated with high SHAP values on the positive side of the X axis. \label{fig:f5}}
\end{figure}

\begin{figure}
\centering
{\includegraphics[width=0.49\textwidth]{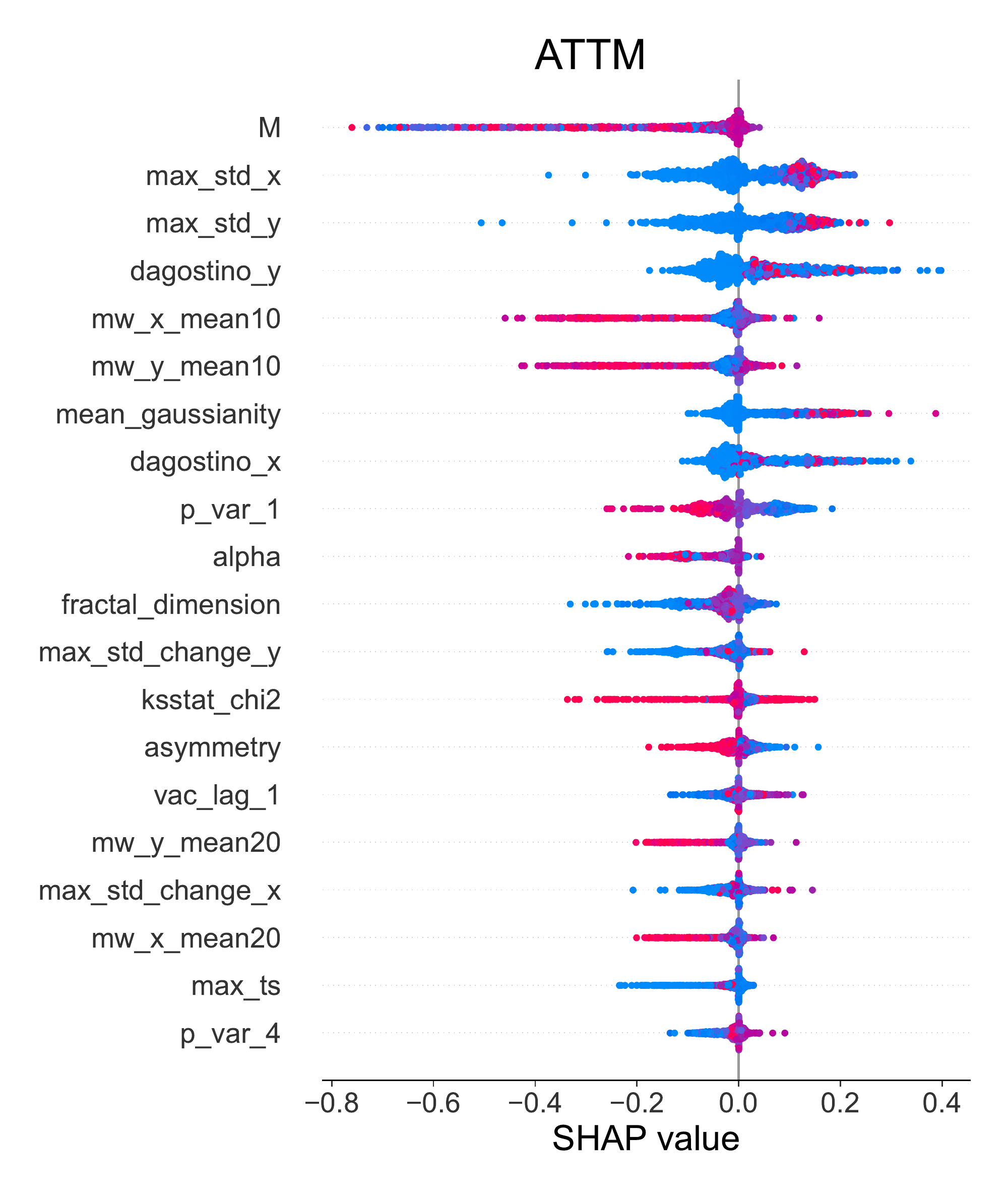}} 
{\includegraphics[width=0.49\textwidth]{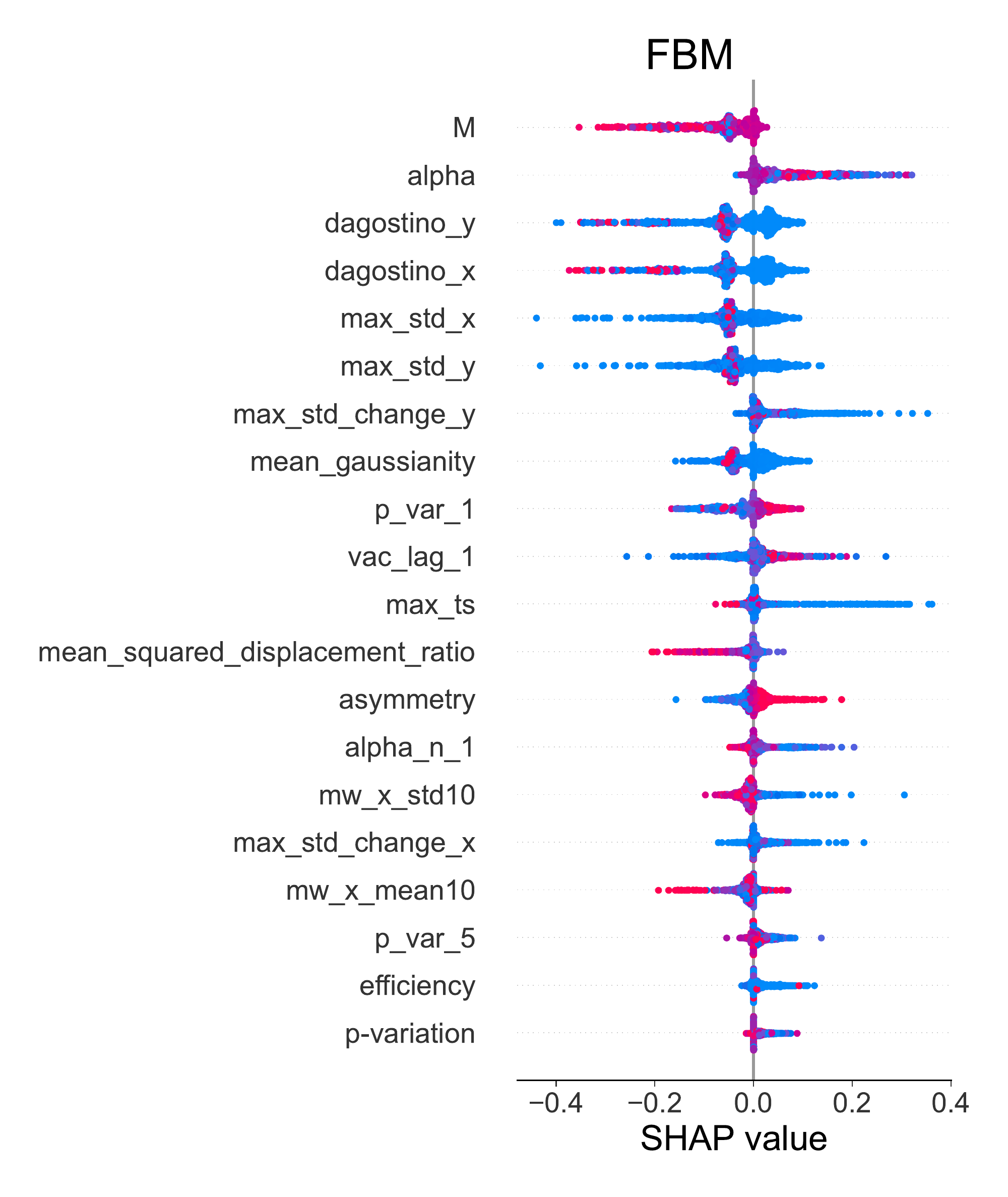}} 
{\includegraphics[width=0.49\textwidth]{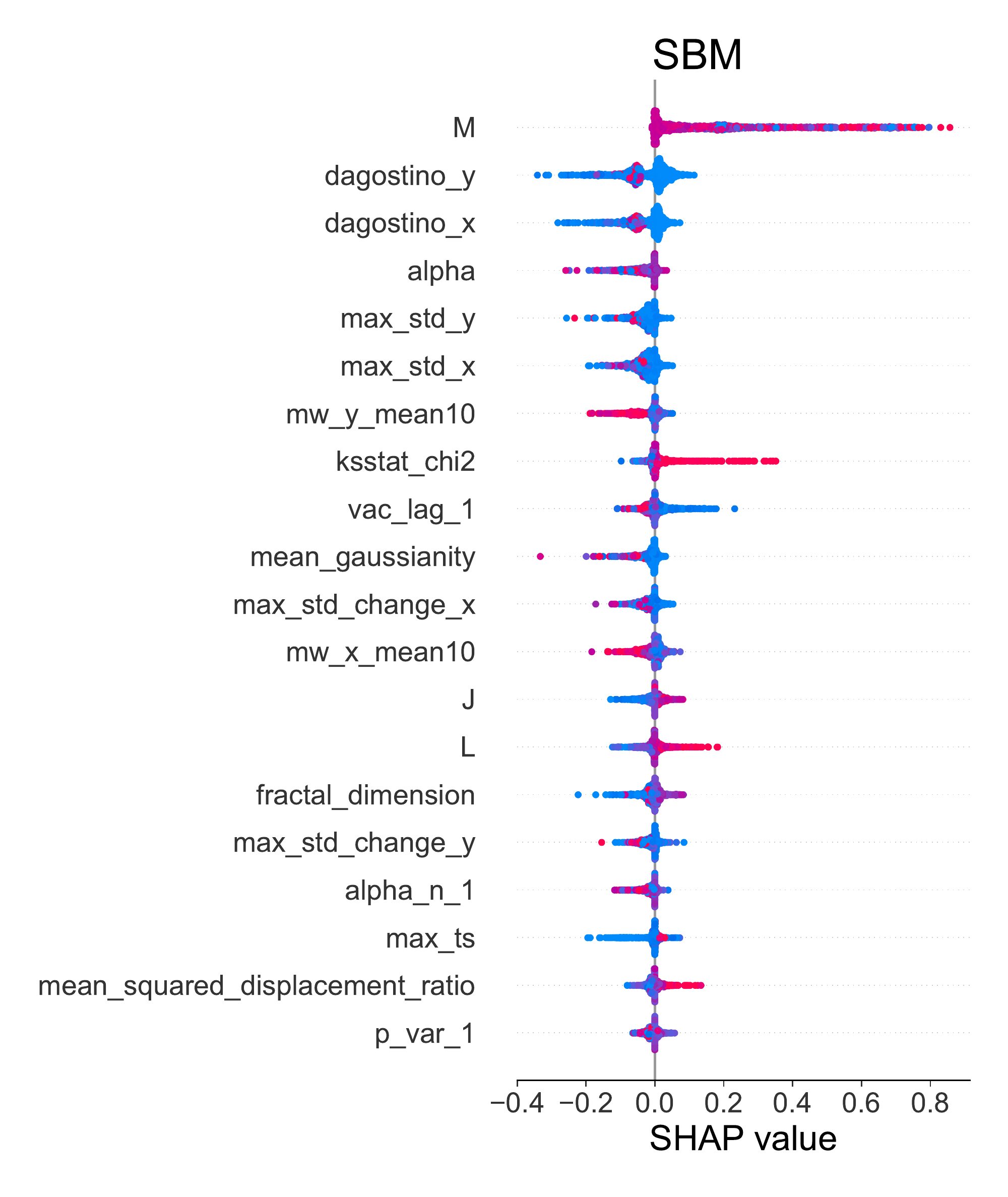}} 
\caption{SHAP values for the 20 most relevant features in the case of ATTM, FBM, and SBM. See caption of Fig.~\ref{fig:f5} for further details. \label{fig:f6}}
\end{figure}

In Figs.~\ref{fig:f5} and~\ref{fig:f6}, we may check how each of the 20 most relevant features impacts the decisions of the classifier for each diffusion model. A dot in the plot corresponds to a single instance of the explanation given by the classifier. The characteristics on the Y axis are sorted by their mean importance. The SHAP values on the horizontal axis reflect the positive or negative influence of each factor on the model prediction. A positive SHAP value increases the likelihood that the sample will be classified as a specific model, whereas a negative SHAP value - that the observation belongs to one of the other classes. Larger values of SHAP (positive or negative) indicate a greater impact on the final prediction. Additionally, the colormap represents the feature values -- red being high and blue being low (the color values are relative). The dependent variable Y is positively correlated with high SHAP values on the positive side of the X axis. In other words, those values have a beneficial effect on prediction. For instance, in the case of CTRW, the larger the value of $mw\_y\_mean\_10$,  the higher SHAP attribution, thus the likelihood of classification as CTRW increases.

As can be seen, the SHAP values for CTRW and LW indicate that our classifier is more likely to distinguish these classes from the rest (some features have a long-tailed distribution of values and the colors indicating positive and negative impact are well separated from each other). In contrast, ATTM, FBM and SBM models reveal a high concentration of samples near 0 for many features. Moreover, the colors are mixed with each other indicating that there is no unique relationship between the SHAP value of a feature and its impact on the prediction. In this case, only a combination of all features can give us a reasonable outcome. In other words, it would be rather impossible to build a statistical test to distinguish these three models that operates with a single feature (among the ones present in our set).

\begin{center}
\begin{table}
\begin{tabular}{|lr|lr|lr|}
\hline\hline
\multicolumn{2}{|l|}{\textbf{ATTM}}                                          & \multicolumn{2}{l|}{\textbf{FBM}}                                           & \multicolumn{2}{l|}{\textbf{SBM}}                                           \\ \hline
\multicolumn{1}{|l|}{model characteristic} & \multicolumn{1}{l|}{shap value} & \multicolumn{1}{l|}{model characteristic} & \multicolumn{1}{l|}{shap value} & \multicolumn{1}{l|}{model characteristic} & \multicolumn{1}{l|}{shap value} \\ \hline
\multicolumn{1}{|l|}{M}                    & 0.13                            & \multicolumn{1}{l|}{M}                    & 0.07                            & \multicolumn{1}{l|}{M}                    & 0.20                            \\ \hline
\multicolumn{1}{|l|}{max\_std\_x}          & 0.08                            & \multicolumn{1}{l|}{alpha}                & 0.06                            & \multicolumn{1}{l|}{dagostino\_y}         & 0.05                            \\ \hline
\multicolumn{1}{|l|}{max\_std\_y}          & 0.08                            & \multicolumn{1}{l|}{dagostino\_y}         & 0.06                            & \multicolumn{1}{l|}{dagostino\_x}         & 0.04                            \\ \hline
\multicolumn{1}{|l|}{dagostino\_y}         & 0.07                            & \multicolumn{1}{l|}{dagostino\_x}         & 0.06                            & \multicolumn{1}{l|}{alpha}                & 0.03                            \\ \hline
\multicolumn{1}{|l|}{mw\_x\_mean10}        & 0.06                            & \multicolumn{1}{l|}{max\_std\_x}          & 0.05                            & \multicolumn{1}{l|}{max\_std\_y}          & 0.03                            \\ \hline
\multicolumn{1}{|l|}{mw\_y\_mean10}        & 0.06                            & \multicolumn{1}{l|}{max\_std\_y}          & 0.05                            & \multicolumn{1}{l|}{max\_std\_x}          & 0.03                            \\ \hline
\multicolumn{1}{|l|}{mean\_gaussianity}    & 0.06                            & \multicolumn{1}{l|}{max\_std\_change\_y}  & 0.03                            & \multicolumn{1}{l|}{mw\_y\_mean10}        & 0.02                            \\ \hline
\multicolumn{1}{|l|}{dagostino\_x}         & 0.06                            & \multicolumn{1}{l|}{mean\_gaussianity}    & 0.03                            & \multicolumn{1}{l|}{ksstat\_chi2}         & 0.02                            \\ \hline
\multicolumn{1}{|l|}{p\_var\_1}            & 0.05                            & \multicolumn{1}{l|}{p\_var\_1}            & 0.03                            & \multicolumn{1}{l|}{vac\_lag\_1}          & 0.02                            \\ \hline
\multicolumn{1}{|l|}{alpha}                & 0.05                            & \multicolumn{1}{l|}{vac\_lag\_1}          & 0.03                            & \multicolumn{1}{l|}{mean\_gaussianity}    & 0.02                            \\ \hline\hline
\multicolumn{2}{|l|}{\textbf{CTRW}}                                          & \multicolumn{2}{l|}{\textbf{LW}}  

& \multicolumn{2}{l|}{\multirow{12}{*}{}}                                  
\\ \cline{1-4}
\multicolumn{1}{|l|}{model characteristic} & \multicolumn{1}{l|}{shap value} & \multicolumn{1}{l|}{model characteristic} & \multicolumn{1}{l|}{shap value} & \multicolumn{2}{l|}{}                                                       \\ \cline{1-4}
\multicolumn{1}{|l|}{mw\_x\_mean10}        & 0.07                            & \multicolumn{1}{l|}{max\_std\_x}          & 0.05                            & \multicolumn{2}{l|}{}                                                       \\ \cline{1-4}
\multicolumn{1}{|l|}{mw\_y\_mean10}        & 0.07                            & \multicolumn{1}{l|}{max\_std\_y}          & 0.05                            & \multicolumn{2}{l|}{}                                                       \\ \cline{1-4}
\multicolumn{1}{|l|}{fractal\_dimension}   & 0.04                            & \multicolumn{1}{l|}{dagostino\_y}         & 0.02                            & \multicolumn{2}{l|}{}                                                       \\ \cline{1-4}
\multicolumn{1}{|l|}{dagostino\_x}         & 0.03                            & \multicolumn{1}{l|}{p\_var\_1}            & 0.02                            & \multicolumn{2}{l|}{}                                                       \\ \cline{1-4}
\multicolumn{1}{|l|}{ksstat\_chi2}         & 0.02                            & \multicolumn{1}{l|}{dagostino\_x}         & 0.02                            & \multicolumn{2}{l|}{}                                                       \\ \cline{1-4}
\multicolumn{1}{|l|}{mw\_x\_mean20}        & 0.02                            & \multicolumn{1}{l|}{alpha}                & 0.02                            & \multicolumn{2}{l|}{}                                                       \\ \cline{1-4}
\multicolumn{1}{|l|}{mw\_y\_mean20}        & 0.02                            & \multicolumn{1}{l|}{vac\_lag\_2}          & 0.01                            & \multicolumn{2}{l|}{}                                                       \\ \cline{1-4}
\multicolumn{1}{|l|}{dagostino\_y}         & 0.02                            & \multicolumn{1}{l|}{max\_std\_change\_y}  & 0.01                            & \multicolumn{2}{l|}{}                                                       \\ \cline{1-4}
\multicolumn{1}{|l|}{mean\_gaussianity}    & 0.02                            & \multicolumn{1}{l|}{max\_std\_change\_x}  & 0.01                            & \multicolumn{2}{l|}{}                                                       \\ \cline{1-4}
\multicolumn{1}{|l|}{p\_var\_1}            & 0.01                            & \multicolumn{1}{l|}{mw\_y\_mean10}        & 0.01                            & \multicolumn{2}{l|}{}                                                       \\ \hline\hline
\end{tabular}
\caption{SHAP values for most important features for different diffusion models in the case of the extended XGB model.\label{tab:featureimportance}}
\end{table}
\end{center}

\section{Conclusion}

In this paper, we have introduced two sets of attributes that may be used for the classification of SPT data with help of feature-based machine learning methods. The first of these sets was proposed as our contribution to Task 2 of the AnDi challenge~\cite{andi2020,munozgil2021objective}. The latter one is the result of our attempt to improve the performance of the classifier presented in the challenge.

The extreme gradient boosting was used as the classification model. It is an ensemble method, which combines multiple decision trees to obtain better predictive performance. Although the deep-learning approach constitutes the state-of-the-art technology for data classification in many domains, we deliberately decided to pick one of the traditional machine learning algorithms due to its superior 
interpretability.

Our original method turned out to be inferior to the top AnDi challenge teams, but still offered a reasonable performance of 73\%. Moreover, further elaboration on the feature set after the challenge allowed us to significantly improve its accuracy by adding some new characteristics and to achieve performance similar to the winning teams (83\% in 2D). Although the final accuracy is already quite good, there is still room for improvement, in particular in the case of short trajectories. Thus, more research into meaningful characteristics is encouraged.

Our results show how crucial the choice of features is for good classification performance. Moreover, we were able to identify the most important among the features for all diffusion types. It does not only contribute to our overall understanding of the decisions made by the classifier. The analysis of feature importance may also be helpful in the selection of meaningful features that can be then used to build simpler classifiers trained on subsets of the attributes.

\begin{acknowledgments}
The work of P.K., H. L.-O., and J.S. was supported by NCN-DFG Beethoven Grant No. 2016/23/G/ST1/04083.
Calculations have been partially carried out using resources provided by Wroc\l{}aw Centre for Networking and Supercomputing (http://wcss.pl).
\end{acknowledgments}

\nocite{*}

\bibliography{references.bib}

\end{document}